\documentclass[aps,superscriptaddress,pra,twocolumn,showpacs,10pt]{revtex4-1}
\usepackage[utf8]{inputenc}
\usepackage[T1]{fontenc}
\usepackage{amsmath}
\usepackage{amsfonts}
\usepackage{amssymb}
\usepackage{color}

\newcommand{\plustimes}{\mathpalette\plustimesinner\relax}
\newcommand{\plustimesinner}[2]{\mathbin{\vphantom{+}\ooalign{$#1+$\cr\hidewidth$#1\times$\hidewidth\cr}}}
\usepackage{graphics,graphicx}
\begin{document}
\title{Optimization of periodic single-photon sources}
\author{Peter Adam}
\email{adam.peter@wigner.mta.hu}
\affiliation{Institute for Solid State Physics and Optics, Wigner Research Centre for Physics, Hungarian Academy of Sciences, H-1525 Budapest, P.O. Box 49, Hungary}
\affiliation{Institute of Physics, University of P\'ecs, H-7624 P\'ecs, Ifj\'us\'ag \'utja 6, Hungary}
\author{Matyas Mechler}
\affiliation{MTA-PTE High-Field Terahertz Research Group, H-7624 P\'ecs, Ifj\'us\'ag \'utja\ 6, Hungary}
\author{Imre Santa}
\affiliation{Institute of Physics, University of P\'ecs, H-7624 P\'ecs, Ifj\'us\'ag \'utja 6, Hungary}
\author{M\'aty\'as Koniorczyk}
\affiliation{Institute of Mathematics and Informatics, University of P\'ecs, H-7624 P\'ecs, Ifj\'us\'ag \'utja 6, Hungary}
\date{\today}
\begin{abstract}
We introduce a theoretical framework which is suitable for the
description of all spatial and time-multiplexed periodic single-photon
sources realized or proposed thus far. Our model takes into account
all possibly relevant loss mechanisms. This statistical analysis of
the known schemes shows that multiplexing systems can be optimized in order to produce maximal single-photon probability for various sets of loss parameters  by the appropriate choice of the number of multiplexed units of spatial multiplexers or multiplexed time intervals and the input mean photon pair number, and reveals the physical reasons of the existence of the optimum. We propose a novel time-multiplexed scheme to be realized in bulk optics, which, according to the present analysis, would have promising performance when experimentally realized. It could provide a single-photon probability of 85\% with a choice of experimental parameters which are feasible according to the experiments known from the literature.
\end{abstract}
\maketitle
\section{Introduction}
As it is prevalently known, single-photon sources are of utmost importance in optical quantum information processing as well as in quantum optics. While many optical quantum information processing schemes ---including linear optical quantum computing \cite{Knill2001, Kok2007}, long-distance quantum key distribution \cite{Gisin2002, Scarani2009} and communication \cite{Duan2001, Sangouard2011}, quantum teleportation \cite{Bennett1993, Bouwmeester1997, Chen2008}, tests of quantum nonlocality \cite{Aspect1981, Merali2011, Koniorczyk2011, Adam2012} or boson sampling processors \cite{Spring2013, Broome2013, Tillmann2013, Crespi2013}--- assume the controlled availability of single photons, in the latter case it can be necessary for the creation of certain nonclassical states of light \cite{Gerry1999, Lund2004, He2009, Adam2010, Lee2012}. In the last fifteen years extensive experimental efforts have been under progress aiming at producing efficient single-photon sources. Deterministic sources can be realized using different kinds of single quantum emitter systems such as quantum dots \cite{Santori2002, Strauf2007, Polyakov2011}, diamond color centers \cite{Beveratos2002a, Gaebel2004, Wu2007}, single atoms \cite{McKeever2004, Hijlkema2007}, ions \cite{Keller2004}, and molecules \cite{Lounis2000,Lettow2010} and also using ensembles of cold atoms \cite{Duan2001, Chou2004}.
Nevertheless, each of these methods has certain issues to overcome \cite{Eisaman2011}, including collection efficiency and repetition rates, or the complexity of experimental setups. It seems that in most of the known such systems, the indistinguishability of the produced photons is not high enough for the majority of the practical applications.

These problems stimulated the construction of heralded single-photon sources (HSPS) based on correlated photon pair generation in nonlinear optical media including spontaneous four-wave mixing (SFWM) in optical fibers and spontaneous parametric down-conversion (SPDC) in bulk crystals and waveguides. The latter process has been proven
to be the most flexible and widespread resource for experiments in quantum information processing because highly indistinguishable single photons in an almost ideal single mode with known polarization can be generated with SPDC systems \cite{Pittman2005,Brida2011,Osorio2013,Ramelow2013}.
Unfortunately, the probabilistic nature of the pair generation complicates the creation of a deterministic, that is, either on-demand or periodic single-photon source based on this system. Though the timing can be easily ensured by pulsed pumping, there remains a finite probability of generating either more than one or no photon pairs during an expected heralding event.

In the literature there are two suggested ways for overcoming this problem and enhancing the single-photon probabilities without increasing the output noise: spatial multiplexing \cite{Migdall2002, Shapiro2007} and time multiplexing \cite{Pittman2002, Jeffrey2004, Mower2011}. In an earlier version of time multiplexing the application of a fiber-photon storage loop or a very high finesse photon storage cavity have been proposed for proper timing \cite{Pittman2002, Jeffrey2004}. In Ref.~\cite{Mower2011} an actively time-multiplexed scheme with a multistage delay line was presented that can be realized on a silicon-on-insulator photonic integrated circuit. Recently, combination of spatial and time multiplexing has also been proposed \cite{Glebov2013}. Thus far only spatial multiplexing has been demonstrated experimentally \cite{Ma2011, Collins2013}.

In this paper we provide a detailed statistical description which is
applicable to all known kinds of multiplexed sources, aiming at the
maximization of single-photon probabilities under realistic
experimental conditions taking into account the possible loss
mechanisms. We analyze these multiplexed systems for various sets of
loss parameters.  Moreover, we propose a novel bulk time-multiplexed
scheme the performance of which can be the best considering
state-of-art experimental technology, comparing to the other known
schemes, according to the analysis.

The paper is organized as follows. In Sec.~\ref{sec:overview} we review the known multiplexed periodic single-photon sources. In Sec.~\ref{sec:nobuti} a novel bulk time-multiplexed scheme is presented. Section~\ref{sec:statdesc} introduces the proposed statistical description which is applicable to all known kinds of multiplexed sources. In Sec.~\ref{sec:analysis} we use the proposed scheme to analyze various kinds of multiplexing schemes, namely an ideal multiplexing system, a spatial multiplexer, storage cavity based multiplexer, and the proposed bulk time multiplexing scheme. Finally, in Sec.~\ref{sec:concl} we summarize our results.

\section{Overview of multiplexed periodic single-photon sources}\label{sec:overview}
A spontaneous parametric down-conversion process generates photon pairs probabilistically. A strong continuous or pulsed laser field, the pump, enters a crystal with second order optical nonlinearity. The interaction with the crystal results in the conversion of some of the pump photons into simultaneously generated photons of lower frequency. While the frequencies are determined by the energy conservation, the wave numbers, and thus the propagation direction of the generated photons is determined by the conservation of momentum, termed as phase matching condition in this context. Altogether there are direction pairs in which there are photons arriving at random instants, but completely correlated in time: if there is a photon in one of the directions (called the signal photon), it is sure that there is a corresponding one propagating to the idler direction (the idler photon) at the same time \cite{Hong1986,Hong1987}. Obviously, spectral filtering has to be applied to select the highly correlated photon pairs in different SPDC sources.

After the filtering the probability of generating $n$ signal/idler
pairs within a measurement time interval $\Delta t$ can be described
with thermal statistics. For weaker spectral filtering the statistics
approaches the Poissonian limit.  Due to this nature of the SPDC
process there is a finite probability of obtaining either no photons
or multiple photon pairs at a time. As a consequence, detecting an idler
photon with a standard single-photon detector, which does not
distinguish multi-photon events from single-photon events, heralds the
presence of the signal photon/photons, thus yielding a heralded
single-photon source far from ideal.

A way of creating a deterministic periodic single-photon source from this probabilistic one is the spatial multiplexing of $N$ single SPDC sources, i.e., multiplexed units, pumped by a pulsed laser, into a single one. In such systems, the input pump power $I$ of the whole multiplexed system is chosen high enough to ensure high probability of obtaining at least one photon pair, while the pump power  $I/N$ of a single SPDC source is low enough that the probability of generating more than one pair in a multiplexed unit can be neglected. Obviously, the period of this multiplexed HSPS is equal to the period of the pulsed laser.

Time multiplexing is another possible way of addressing this problem. Time multiplexing schemes can be divided into storage cavity-based and cascade delay-based schemes. 
The common feature of these techniques is that in order to achieve a periodic source of period $T_{p}$, we choose an observation time $T<T_{p}$ for which we expect the arrival of at least one signal-idler pair. We divide this observation time to smaller time windows of length $\Delta t$. If the system is pumped by a pulsed laser, $\Delta t$ will be the pumping period, while for continuous pumping the detector of the idler photons is active for such periods. If an idler photon is detected at a given time window, its signal counterpart is delayed to such an extent that finally it leaves at the end of the time $T$. Thus these time windows are the counterparts of multiplexed units in such schemes.

\begin{figure}[tb]
\includegraphics[width=\columnwidth]{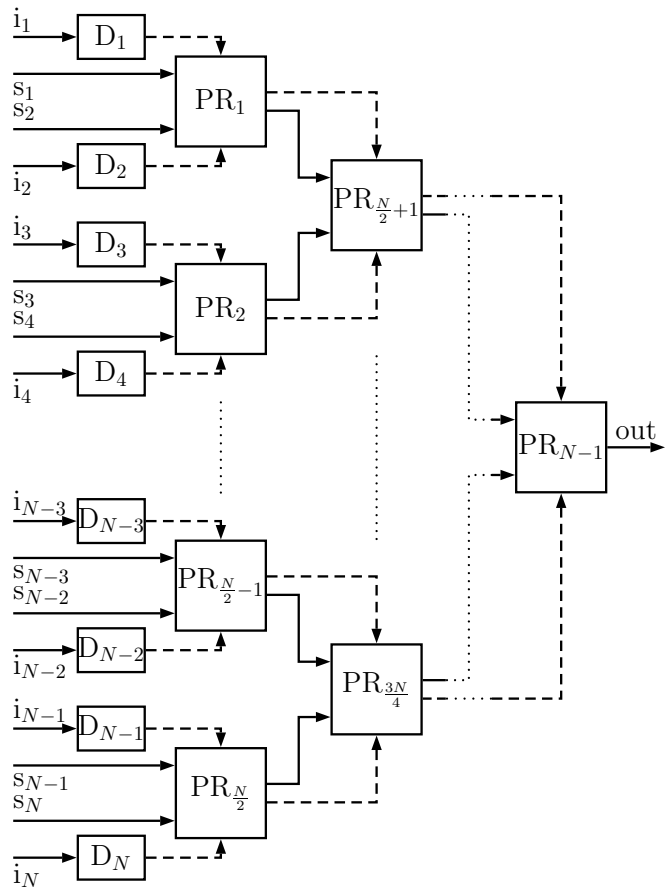}
\caption{\label{fig:ter:model}Schematics of a spatially multiplexed periodic single-photon source. PR$_j$: $j$th photon router, the D$_j$-s are detectors, i$_j$ and s$_j$ are the idler and signal arms of the $j$th SPDC source. Dashed lines represent electronic control lines.}
\end{figure}

Let us now consider the operation of spatially multiplexing schemes in more detail. The key ingredients of the scheme are the photon routers. These devices are electronically controlled. They have multiple input ports and a single output. The control signal determines which single input port is directed to the output, while the others are closed. 
Consider now $N$ sources of photon pairs, preferably pumped by the same strong pulsed laser. As the probability of the generation of a signal-idler pair is low, it is likely that only one of the sources will emit a pair. The detection of the idler can show which of the sources provided the signal photon. If all the signal modes are fed into a router, then the router sends the generated photon into its single output port. Thereby it is certain that we do have a signal photon, and it will be directed to the output. For the case when there were multiple signal-idler pairs generated in the same period, a priority logic should be implemented in the router control to prefer only one of the signal photons. Of course the time required by the operation of the switching should be compensated with a properly designed delay line in the signal port. In this way a router can merge multiple SPDC sources into a single one.

For we intend to study practical issues of such a scheme such as loss and efficiency, we have to take into account that the prevalently available routers have only two input ports. Thus a single router is capable of merging two SPDC sources into a single, more efficient source. This leads us to the cascaded scheme depicted in Fig.~\ref{fig:ter:model}, which is implemented in the known experiments. Here we have pairs of SPDC sources (all pumped by the same laser) merged at the first level. At the next level the outputs of the pairs are arranged into pairs, and the detector signal is also forwarded to control the routers of the next level. Thus the pairs of the first level are merged pairwise. Finally there will be a single output only. Of course the already mentioned priority logics as well as the delays should be designed appropriately. Using the notation of Fig.~\ref{fig:ter:model}, such a system needs $N=2^m$ input photon pairs and, accordingly, $N$ detectors and $N-1$ photon routers. We remark here that even though these schemes were first demonstrated in bulk optics, due to the large number of required elements it is likely that it would be scalable in integrated optical applications only.

\begin{figure}[tb]
\includegraphics[width=\columnwidth]{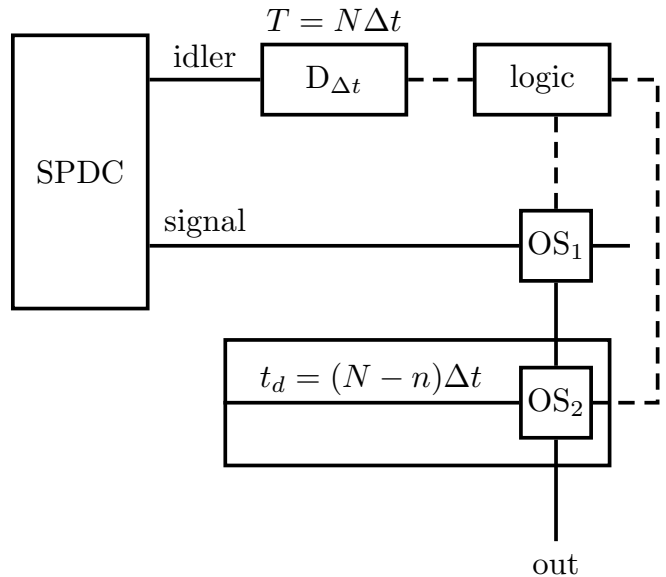}
\caption{\label{fig:tgy:model}Schematics of a storage cavity-based periodic photon source. SPDC is a spontaneous parametric down-conversion source yielding twin photon pairs; OS$_1$ and OS$_2$ are optical switches. $t_d$ is the delay introduced by the cavity if the idler photon was detected in the $n$-th time window of
length $\Delta t$ within the period of $T=N\,\Delta t$.}
\end{figure}
Now let us turn our attention to time-multiplexed schemes. Figure~\ref{fig:tgy:model} shows the arrangement for a storage cavity-based time-multiplexed scheme. Again, if an idler photon is detected by the detector D$_{\Delta t}$, the detector signal triggers a logic which controls the switches OS$_1$ and OS$_2$ to direct the heralded photon to a storage cavity. At the end of the observation time $T$ the logic controls the second switch, OS$_2$ to release the photon, thereby ensuring the appropriate release time of the photon. If the photon was detected in the $n$-th time window of length $\Delta t$ within the period of $T=N\,\Delta t$, the storage cavity introduces a delay of length $t_d=(N-n)\Delta t$. Were there be more photons generated during the time $T$, only the first one will be used. This is the counterpart of the priority logic in this scheme. In case of continuous pumping the inaccuracy of the time of the idler detection is $\Delta t$, thus the jitter of this method is also $\Delta t$. In case of pulsed pumping the idler detection can be more accurate.

The third method analyzed in this article is the cascade delay-based time multiplexing scheme. This was proposed in the context of integrated optics \cite{Mower2011}. In the present paper, however, we propose a new version of it, so we discuss its details in the next section.
\section{A novel bulk time-multiplexed scheme}\label{sec:nobuti}
\begin{figure}[tb]
\includegraphics[width=\columnwidth]{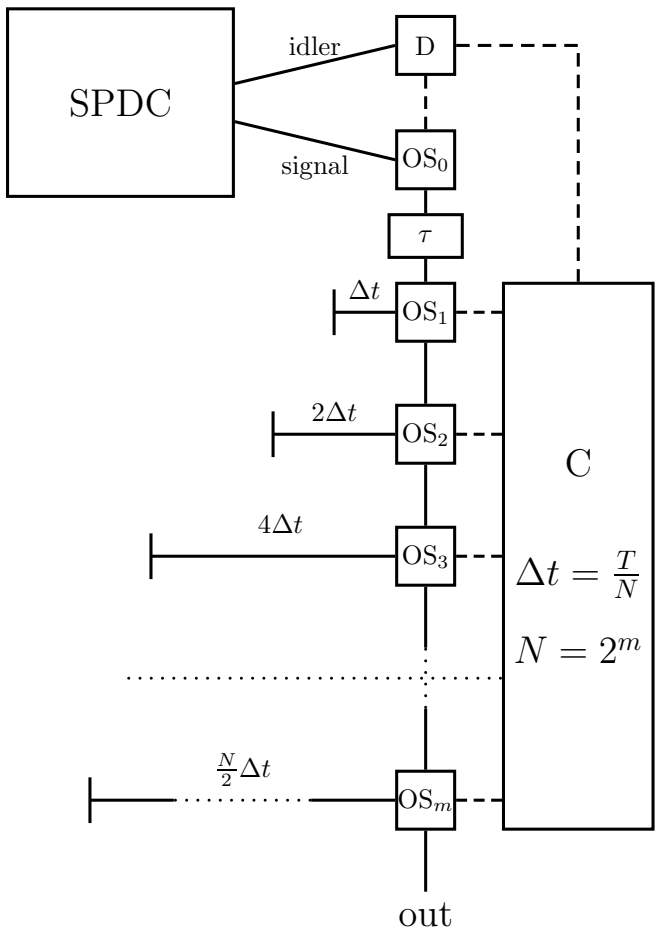}
\caption{\label{fig:ido:model}Periodic photon source. SPDC is a spontaneous parametric down-conversion source yielding twin photon pairs; D: detector unit detecting idler photons and switching OS$_0$; OS$_0$ is an optical switch with a gate width $\Delta t$ that selects the twin signal photon; C is the controller unit setting the optical switches OS$_i$ ($i=1,\dots,m$) ensuring proper delay of the twin photon; the delay $\tau$ is needed for the proper operation of the controller C.}
\end{figure}
\begin{figure}[tb]
\includegraphics[width=\columnwidth]{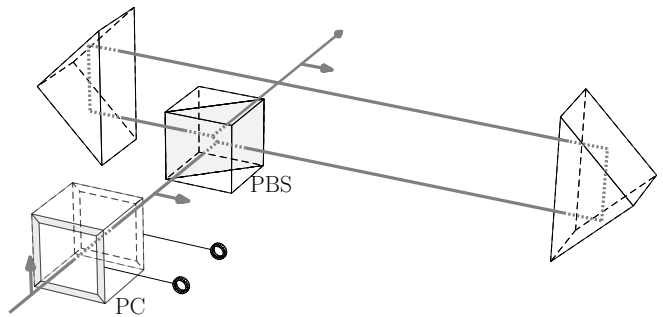}
\caption{\label{fig:ido:egyseg}Schematic figure of a single delay unit. PC: Pockels cell, PBS: polarizing beam splitter. An initially vertically polarized photon arrives at the Pockels cell. By changing its polarization, the photon can be forced to use the delay line.}
\end{figure}
Here we introduce our suggested setup for a time-multiplexed scheme in bulk optics. The scheme is depicted in Fig.~\ref{fig:ido:model}. The idler part of the photon pairs emerging from the continuous or pulsed SPDC source are detected for a time interval of length $T$. Let the mean number of photon pairs arriving in a duration $T$ be $\lambda$. Having detected an idler photon by the detector unit D, its signal pair is directed by the optical switch OS$_0$ to a delay system which introduces such a delay as if the signal photon were to arrive at time $T$. At a time $\Delta t$ after the detection of the idler photon the system shuts, that is, only the signal photons generated in this time window may enter the delay system. We assume that $T=N\cdot\Delta t$, where $N=2^m$, $m$ being an integer. This is due to the discrete nature of the multi-step delay system assumed to be realized in the experiment: there are $m$ switchable delay units (branches) realizing delays of $1\Delta t, 2\Delta t, 4\Delta t,\dots,k\cdot \Delta t$, where $k=N/2$. Hence, if all the $m$ delay units are turned on, the achieved delay is $(N-1)\Delta t$. Were the first photon to arrive in the $n$-th time window, a delay of $(N-n)\Delta t$ has to be applied. In the described delay system it means that we should only apply those delay units that correspond to the digits 1 in the binary numerical representation of $N-n$. In the scheme in Fig.~\ref{fig:ido:model} this is achieved by the use of the appropriate optical switches activated by the control unit.  Let the delay required by this unit to ensure the appropriate delay configuration be denoted by $\tau$. We remark that in the case of pulsed pumping source the pumping period have to be chosen to be equal to $\Delta t$. After the time the signal photon leaves the delay system, the whole procedure of the detection, of time length $T$ is restarted. The described process results in a photon source of period $T_p$. The minimum practically achievable period of such a photon source is
\begin{equation}
\label{eq:minTp}
T_{p,\rm min}=\max(T+\tau+\tau_d,T+\tau_0),
\end{equation}
where $\tau_d$ is the minimal time for the signal photon to pass the delay system without any activated delay unit, whereas $\tau_0$ is the dead time of the detector: the time required by the detector to register the next detection event. We note that in case of continuous pumping the resulting jitter of this scheme is also $\Delta t$.

The switchable delay units may be realized in various ways. In Fig.~\ref{fig:ido:egyseg} we suggest a possible realization of such a unit using a Pockels cell PC, a polarizing beam-splitter PBS, and two prisms, arranged as ``periscopes''. Assume the polarization of the incoming photon to be linear (horizontal or vertical) and to be known in advance. Depending on whether the delay unit should be used or not, the PC changes or keeps the polarization of the photon. The next optical element, the PBS causes S-polarized photon to be reflected at right angle, while a P-polarized photon to be transmitted. The delay is implemented when the photon was S polarized and thus reflected. In this case it enters the double periscope system. The first periscope elevates the reflected photon into a plane perpendicular to the original propagation direction, and makes it propagate backwards along a direction parallel to the one incident to the periscope. The second periscope lowers the photon to the original plane and directs it to the other side of the PBS.  Finally, the PBS makes the photon to return to the original propagation direction.

As an estimate for the particular parameters of the scheme, the minimal feasible time window can be considered to be $\Delta t_{\rm min}=100$~ps. This value is determined by the spatial extent of the delay unit corresponding to a delay of $\Delta t$ and the gate width of the optical switch OS$_0$. The minimal control time of the presented multiplexer is around $\tau_{\rm min}=30$~ns. Hence the currently achievable minimal period of such a source, assuming e.g.\ $m=9$ delay units, according to Eq.~\eqref{eq:minTp} should be around $T_{p,\rm min}\approx 80$~ns.

\section{A framework for the statistical description}\label{sec:statdesc}
In this section we present a common theoretical framework describing all the multiplexed periodic single-photon sources presented in the previous two sections. Assume that the $n$-th time window or multiplexed unit (either of these will be termed as ``unit'' throughout this Section) adds $j$ signal photons to the multiplexing system with probability $P_n^{(j)}$ independently of $n$. (For $j=0$ it is the probability of not a single arriving photon.) The probability of obtaining $i$ photons altogether in a period of the output signal of any of the studied multiplexing systems in general reads
\begin{small}
\begin{eqnarray}
P_0&=&\left(P_n^{(0)}\right)^N+\sum_{n=1}^N\left(P_n^{(0)}\right)^{n-1}\sum_{j=1}^{\infty}\left[\binom{j}{0}P_n^{(j)}V_n^0(1-V_n)^j\right]\nonumber\\
P_i&=&\sum_{j=i}^{\infty}\sum_{n=1}^N\binom{j}{i} \left(P_n^{(0)}\right)^{n-1}P_n^{(j)}V_n^i(1-V_n)^{j-i},\,i\ge1 .\label{eq:gen:Pout}
\end{eqnarray}
\end{small}
In these expressions, $V_n$ is the probability that a signal photon generated in the $n$-th unit reaches the output, that is, it was not lost in the multiplexing system. The term $\left(P_n^{(0)}\right)^N$ describes the case when there are not any photons detected in either of the $N$ units. The second term in the formula of $P_0$ is the joint probability of detecting an idler photon in the $n$-th unit and all the $j$ signal photons coming from this unit are lost meanwhile. Correspondingly, $P_i$ is the joint probability of detecting an idler photon in the $n$-th unit while there are $j$ signal photons arriving from this unit into the system, $i$ of them are transmitted, and $j-i$ are lost. 

Assuming standard not photon-number resolving detectors of efficiency $V_D$, the probabilities $P_n^{(0)}$ and $P_n^{(j)}$ in~\eqref{eq:gen:Pout} can be obtained as
\begin{eqnarray}
{P_n^{(0)}}&=&\sum_{k=0}^{\infty}\binom{k}{0}{P_n^{(k)}}' V_D^0(1-V_D)^k,\nonumber\\
{P_n^{(j)}}&=&{P_n^{(j)}}'\sum_{k=0}^{j-1}\binom{j}{j-k}{V_D}^{j-k}\,{\left(1-V_D\right) }^{k}\,,\label{eq:Pin:det}
\end{eqnarray}
where ${P_n^{(k)}}'$ is the probability of arriving $k$ photon pairs at the multiplexing system in a multiplexed unit. In the case of an SPDC photon source the probabilities ${P_n^{(k)}}'$ in the above expressions can be described by a Poissonian distribution:
\begin{equation}
{P_n^{(k)}}'=\frac{(\lambda/N)^k}{k!}\exp\left(-\frac{\lambda}{N}\right),
\end{equation}
where $\lambda$ is the mean number of photon pairs arriving in a duration $T$ for time multiplexing system, while in the case of spatial multiplexing it is the mean total number of photon pairs arriving at the input ports of the whole multiplexing system. The expression of $P_n^{(0)}$ in Eq.~\eqref{eq:Pin:det} describes the joint probability of arriving any number $k$ of idler photons at the detector and none of them being detected. The expression of $P_n^{(j)}$ describes the case when $j$ idler photon arrives at the detector and the detector clicks (caused by any number of them). The detection results in adding $j$ photons to the multiplexing system. We note that Eqs.~\eqref{eq:gen:Pout} and \eqref{eq:Pin:det} are valid for thermal distribution, too. It is easy to verify that the probabilities in Eq.~\eqref{eq:Pin:det}, as well as the probabilities $P_i$ in Eq.~\eqref{eq:gen:Pout} sum up to 1 as appropriate.

For a detailed analysis of the described systems and the calculation of their properties it is necessary to take losses into account. This we shall describe by a transmission coefficient in the theoretical framework applicable for all the studied systems, albeit its actual form will depend on the particular scheme.

In our proposed time-multiplexed system there are four kinds of losses which may arise. The signal photon may be absorbed or scattered on its way through the medium of the delay system. This we describe by the transmission probability $V_t$ relating to the propagation through the whole medium, that is, the medium of the longest delay. An additional loss due to the specific elements of delay units can arise if a delay unit or branch is either used or not. Let the respective transmission probabilities be denoted by $V_r$ and $V_{r,0}$. These losses originate mainly from the reflection and transmission efficiencies of the polarizing beam splitter of a single delay unit in Fig.~\ref{fig:ido:egyseg}. Assume that the first idler photon is detected in the $n$-th time window and the corresponding signal photon has to be delayed for $(N-n)\Delta t $, and the number of delay branches is $m$. In this case the total probability of transmission will read
\begin{equation}
V_n=V_r^sV_{r,0}^{m-s}V_t^{(N-n)/N}V_b,\label{eq:ido:veszt}
\end{equation}
where $s$ is the Hamming weight of $N-n$ (the number of ones in its binary representation). The coefficient $V_b$ is the generic transmission coefficient independent of the $n$-th time window, which may be due to, e.g. the loss of the optical switch OS$_0$ controlling the path of the signal photon, etc.

In spatial multiplexing systems optical routers are applied. In the cascaded system with $N=2^m$ spatial sources a photon originating from any of the units passes $m$ routers. Hence, the transmission coefficient reads
\begin{equation}
V_n=V_R^{\log_2N}V_b,\label{eq:ter:veszt}
\end{equation}
where  $V_R$ stands for the transmission coefficient of a single router.
In case of cavity-based multiplexing the transmission coefficient reads
\begin{equation}
V_n=V_c^{N-n}V_b,\label{eq:tgy:veszt}
\end{equation}
where $V_c$ is the transmission coefficient of the storage cavity in
case of a single round trip.
\section{Optimal single-photon sources}\label{sec:analysis}
In this Section we present our results regarding the optimization of single-photon probability in various experimental settings obtained by using the statistical model of Sec.~\ref{sec:statdesc}. First we study an ideal multiplexer in general. Then we take into account the
specialties of each discussed scheme.
\subsection{Ideal multiplexers}
\begin{figure}[tb]
\includegraphics[width=\columnwidth]{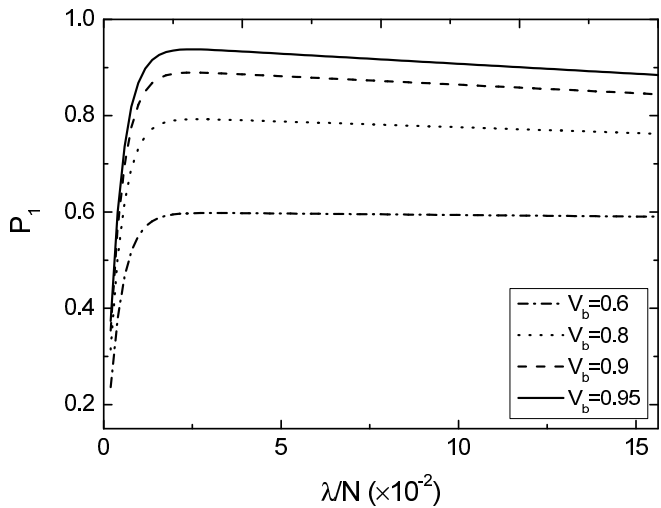}
\caption{\label{fig:konst:P1LpN}The single-photon probability $P_1$ plotted against the mean photon pair number per multiplexed unit $\lambda/N$ for various generic transmission coefficients $V_b$, considering $N=256$ multiplexed units for an ideal multiplexer, and assuming ideal detectors ($V_D=1$).}
\end{figure}
\begin{figure}[tb]
\includegraphics[width=\columnwidth]{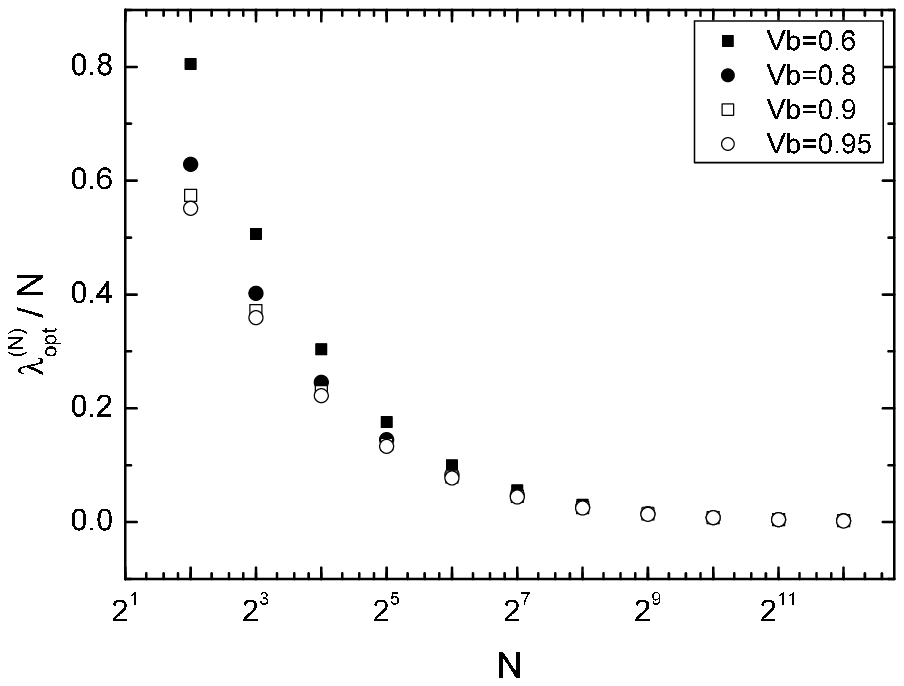}
\caption{\label{fig:konst:LpNN}The optimal choice  $\lambda_{\rm opt}^{(N)}/N$ of the mean photon pair number per multiplexed unit as a function of the number of multiplexed units $N$ on semi-logarithmic scale for different generic transmission coefficients $V_b$ for an ideal multiplexer, and assuming ideal detectors ($V_D=1$).}
\end{figure}
\begin{figure}[tb]
\includegraphics[width=\columnwidth]{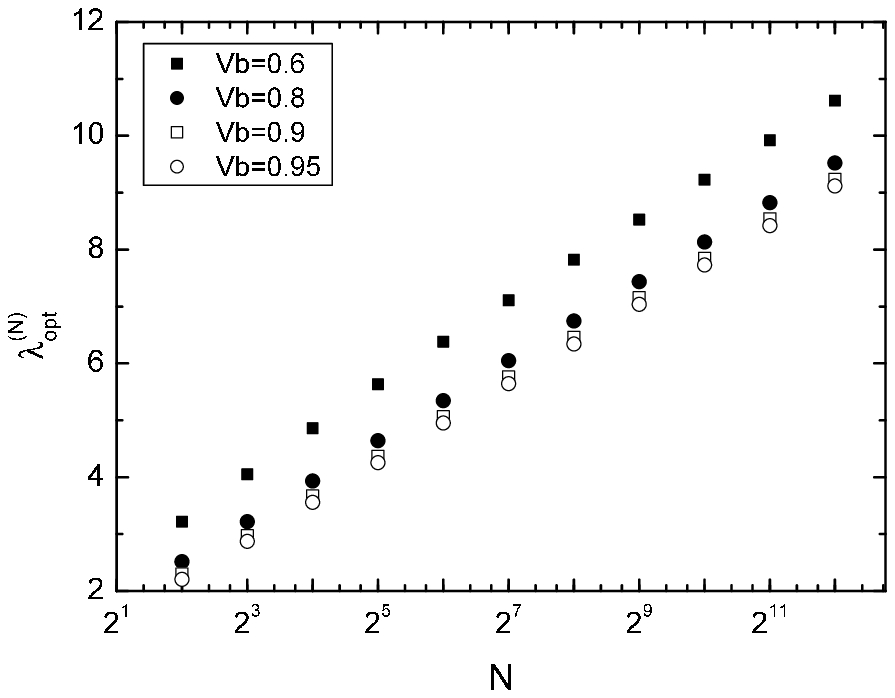}
\caption{\label{fig:konst:lamN}The optimal choice  $\lambda_{\rm opt}^{(N)}$ of the mean photon pair number as a function of the number of multiplexed units $N$ on semi-logarithmic scale for different generic transmission coefficients $V_b$  for an ideal multiplexer, and assuming ideal detectors ($V_D=1$).}
\end{figure}
\begin{figure}[tb]
\includegraphics[width=\columnwidth]{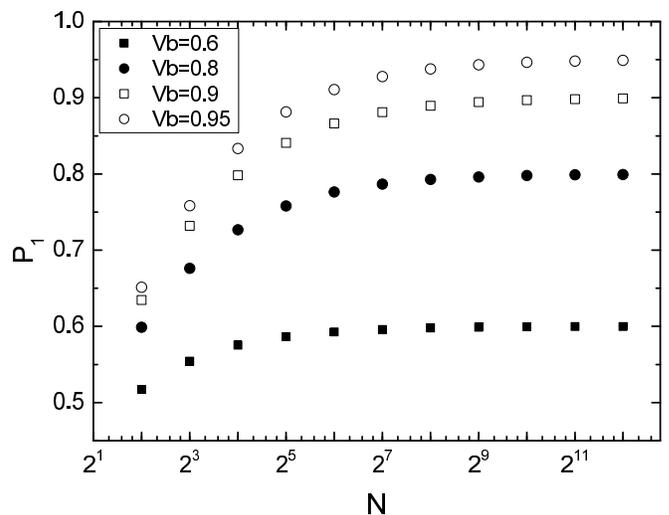}
\caption{\label{fig:konst:P1N}The achievable maximal single-photon probability $P_1$ at the optimal choice $\lambda_{\rm opt}^{(N)}$ of the mean photon pair number as a function of the number $N$ of multiplexed units on semi-logarithmic scale for different generic transmission coefficients $V_b$ for an ideal multiplexer, and assuming ideal detectors ($V_D=1$).}
\end{figure}

Let us first investigate an idealized case when the loss in the multiplexing system is independent of the number of multiplexed units so it is a constant. In case of the arrangements discussed in the previous section it means that each transmission coefficient is equal to one, except for the $V_b$ generic transmission coefficient in Eqs.~\eqref{eq:ido:veszt}, \eqref{eq:ter:veszt}, and \eqref{eq:tgy:veszt}. In Fig.~\ref{fig:konst:P1LpN} the single-photon probability $P_1$ is plotted against $\lambda/N$, the mean photon pair number per multiplexed unit, for $N=256$ units and various values of $V_b$ as a parameter. Let us note that the figure would be alike for an arbitrary $N$ number of units. It appears that for given values of $N$ and $V_b$ the probability $P_1(\lambda)$ has a maximum, thus there exists an optimal choice $\lambda_{\rm opt}^{(N)}$ of the mean photon pair number for which the maximal probability of single photons is obtained. The physical reason is clear: for low mean photon pair numbers ($\lambda\to 0$) the probability of obtaining no photons will increase, while a higher mean photon pair number makes the appearance of multiple photons in a single time window more likely.

In Fig.~\ref{fig:konst:LpNN} we can see the dependence of the optimal choice $\lambda_{\rm opt}^{(N)}/N$ as a function of the number of multiplexed units $N$, for various $V_b$ generic losses and still for ideal detectors ($V_D=1$), while Fig.~\ref{fig:konst:lamN} shows the dependence of $\lambda_{\rm opt}^{(N)}$ (not divided by $N$) in the same way. We note that logarithmic scale for $N$ is used for ensuring comparability with the same plots of other multiplexing schemes discussed afterwards. The two figures illustrate the essence of the necessary considerations for multiplexing: $\lambda_{\rm opt}^{(N)}$ increases with the number of units, as in this case the probability of obtaining no photon pairs decreases. Meanwhile $\lambda_{\rm opt}^{(N)}/N$ decreases, hence, there is less chance for multiple photons at the output.

In Fig.~\ref{fig:konst:P1N} one can see the achievable maximal single-photon probability $P_1$ at the optimal choice $\lambda_{\rm opt}^{(N)}$ of the mean photon pair number as a function of the number $N$ of multiplexed units. The highest $P_1$ is achievable with $N\to \infty$. For a given transmission coefficient $V_b$, the maximal probability $P_{1,\rm max}$ is just equal to $V_b$. Let us note that $P_1$ gets close to its maximum already for a relatively small number of units. For $V_b=0.9$ and $N=256$, for instance, the maximal probability is $P_1=0.8895$ at the mean photon pair number $\lambda_{\rm opt}^{(N)}=6.46$, or with respect to a single
unit, $\lambda_{\rm opt}^{(N)}/N=0.025$.

After discussing the idealized case in general, now let us discuss the described schemes in the presence of loss, which makes the behavior dependent on the particular arrangement.
\subsection{Spatial multiplexers}
\begin{figure}[tb]
\includegraphics[width=\columnwidth]{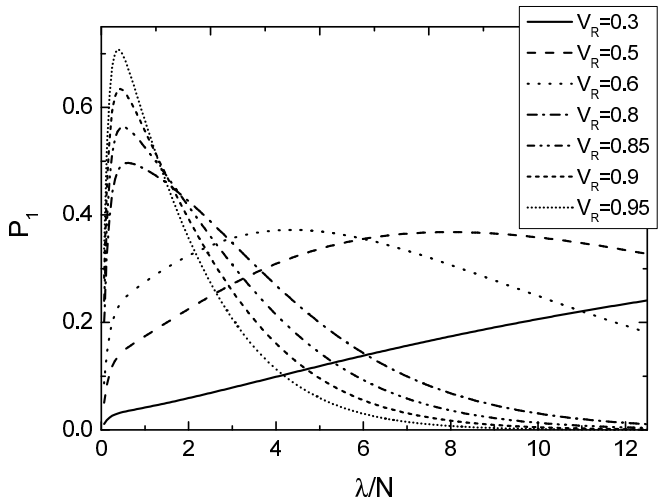}
\caption{\label{fig:ter:P1L}
The single-photon probability $P_1$ plotted against the mean photon pair number per multiplexed unit $\lambda/N$, for various router transmission coefficients $V_R$, considering $N=8$ multiplexed units for a spatial multiplexer, assuming ideal detectors and no generic losses ($V_D=V_b=1$).}
\end{figure}
\begin{figure}[tb]
\includegraphics[width=\columnwidth]{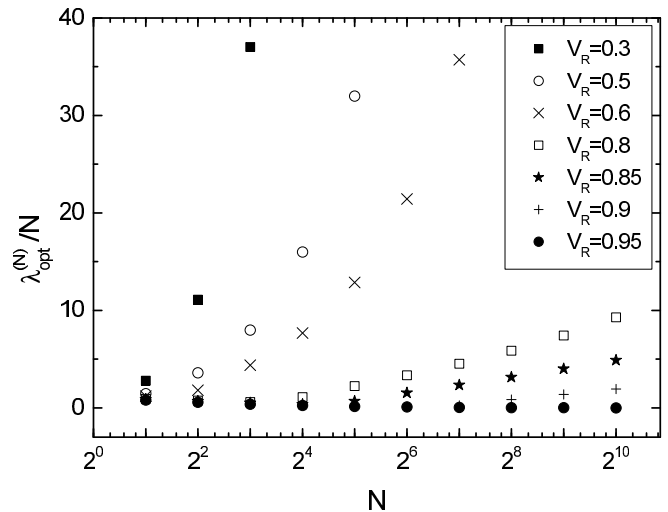}
\caption{\label{fig:ter:LpNN}
The optimal choice  $\lambda_{\rm opt}^{(N)}/N$ of the mean photon pair number per multiplexed unit as a function of the number of multiplexed units $N$ on semi-logarithmic scale for different router transmission coefficients $V_R$ for a spatial multiplexer, assuming ideal detectors and no generic losses ($V_D=V_b=1$).}
\end{figure}
\begin{figure}[tb]
\includegraphics[width=\columnwidth]{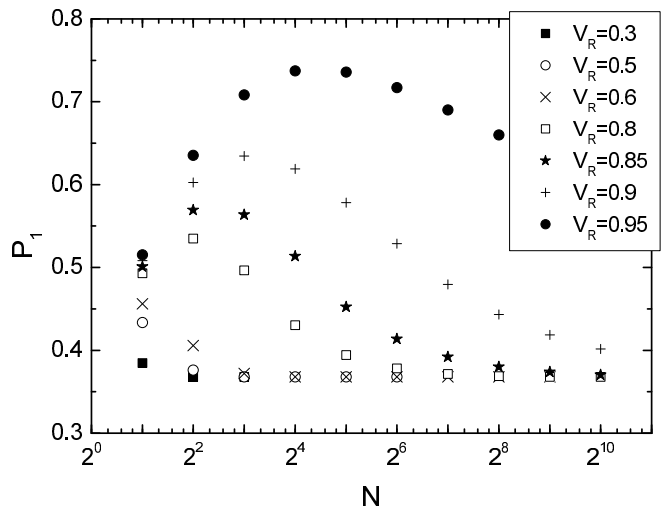}
\caption{\label{fig:ter:P1N}
The achievable maximal single-photon probability $P_1$ at the mean photon pair number $\lambda_{\rm  opt}^{(N)}$ as a function of the number $N$ of multiplexed units
on a semi-logarithmic scale for different router transmission coefficients $V_R$ for a spatial multiplexer, and assuming ideal detectors and no generic losses ($V_D=V_b=1$).}
\end{figure}
Now we analyze spatial multiplexing, in the case of which losses are
described by Eq.~\eqref{eq:ter:veszt}. In Fig.~\ref{fig:ter:P1L} the
single-photon probability $P_1$ is plotted against $\lambda/N$ for $N=8$ units and different values $V_R$ of the transmission coefficient of the
multiplexing router. Here we assume that there are neither generic
nor detector losses in the system, that is, $V_D=V_b=1$. In fact a
value of $V_R=0.3$ is the best feasible value in current integrated
optics, and the theoretical upper bound \cite{Ljunggren2005,Ma2011} which may be
feasible in any kind of such a system is
$V_R=0.95$. Figure~\ref{fig:ter:P1L} shows that the single-photon
probability $P_1$ is much more sensitive to the change of the mean
photon pair number than the ideal arrangement discussed in the
previous subsection. The value $\lambda_{\rm opt}^{(N)}/N$ at which
the maximum of $P_1$ is reached grows with the growth of the losses,
for values below $0.6$ of the coefficient $V_R$ to a higher and higher
extent. The reason is that the mean photon pair number growth compensates for the higher
losses.

Figure~\ref{fig:ter:LpNN} shows the values of $\lambda_{\rm
  opt}^{(N)}/N$ corresponding to the maximal values of $P_1$ as a
function of the number $N$ of multiplexed units on a semi-logarithmic
scale. Note that the behavior of these functions differs from the ones presented for an ideal multiplexer in Fig.~\ref{fig:konst:LpNN} which is the counterpart of this figure.
With the growth of the number of multiplexed units (or, otherwise speaking, the growth of the number $m=\log_2N$ of cascading router levels), the required optimal mean photon pair number for a single unit decreases initially, but after a given number of routers it starts to grow, hence it has a minimum. For $V_R=0.95$, this minimum is at $N^{\min}=8192$, for $V_R=0.9$ it is at $N^{\min}=64$, for $V_R=0.85$, $N^{\min}=16$, and for $V_R=0.8$ it is at $N^{\min}=8$. For a value of $V_R=0.6$ there is no such extremum, the required $\lambda_{\rm opt}^{(N)}/N$ simply grows with $N$ (or $m$). This compensates for the growth of loss as described by Eq.~\eqref{eq:ter:veszt}.

Figure~\ref{fig:ter:P1N} shows the maximal values of $P_1$ as a function of $N$. It can be seen that in contrast to the case of the ideal multiplexer (c.f.\ Fig.~\ref{fig:konst:P1N}), this has a maximum at a given number of multiplexed units. This is the absolute maximum of the single-photon probability $P_{1,\rm max}$ which can be achieved by spatial multiplexing with the optimal choice of the mean photon pair number and the number of multiplexed units (or router levels) subject to the given
losses. The existence of this maximum is due the fact that the growth
of the cascaded levels significantly increases losses, which
deteriorates the benefit of multiplexing. It appears that if
$N\to\infty$ ($m\to\infty$), for any value of $V_R<1$, the
single-photon probability $P_1$ tends to $\exp(-1)$, which is just
the achievable maximum without multiplexing. For the mean photon pair
number per multiplexed units corresponding to this limit,
$\lambda_{\rm opt}^{(N)}/N\to V_R^{-m}$ holds. All these can be easily
derived from Eqs.~\eqref{eq:gen:Pout} and \eqref{eq:ter:veszt}.

\begin{table*}
\caption{\label{tab:ter:ideallams}
Maximal single-photon probabilities $P_{1,\rm max}$ and the required number of router levels $m_{\rm opt}=\log_2N_{\rm opt}$ and $\lambda_{\rm opt}$ at which they can be achieved, calculated for different $V_R$ multiplexing router transmissions and three different values of the detector loss $V_D$. The corresponding zero-photon probabilities $P_{0,\rm max}$ are also given.}
\begin{ruledtabular}
\begin{tabular}{ll|rrll|rrll|rrll}
&&\multicolumn{4}{c|}{$V_D=1.0$}&\multicolumn{4}{c|}{$V_D=0.9$}&\multicolumn{4}{c}{$V_D=0.2$}\\
No.&$V_R$&$m_{\rm opt}$&$\lambda_{\rm opt}$&$P_{1,\rm max}$&$P_{0,\rm max}$&
$m_{\rm opt}$&$\lambda_{\rm opt}$&$P_{1,\rm max}$&$P_{0,\rm max}$&
$m_{\rm opt}$&$\lambda_{\rm opt}$&$P_{1,\rm max}$&$P_{0,\rm max}$\\\hline
1.&0.3&	1&5.60&0.385&0.397&
		1&5.63&0.385&0.392&
		2&43.00&0.369&0.369\\
2.&0.5&	1&2.90&0.434&0.364&
		1&3.03&0.423&0.356&
		3&59.38&0.371&0.372\\
3.&0.6&	1&2.41&0.456&0.331&
		1&2.52&0.439&0.330&
		3&30.18&0.379&0.375\\
4.&0.8&	2&3.03&0.535&0.312&
		2&3.19&0.521&0.308&
		4&20.50&0.422&0.378\\
5.&0.85&2&2.73&0.569&0.264&
		3&4.20&0.556&0.336&
		5&25.25&0.449&0.409\\
6.&0.9&	3&3.44&0.635&0.255&
		3&3.63&0.621&0.254&
		5&16.93&0.515&0.332\\
7.&0.95&4&3.89&0.737&0.185&
		5&5.04&0.729&0.220&
		7&21.27&0.648&0.282
\end{tabular}
\end{ruledtabular}
\end{table*}
In Tab.~\ref{tab:ter:ideallams} we have listed maximal single-photon
probabilities $P_{1,\rm max}$ and the required number of router levels
$m_{\rm opt}=\log_2N_{\rm opt}$ and $\lambda_{\rm opt}$, calculated
for different $V_R$ multiplexing router transmissions and three
different values of the detector loss $V_D$. The corresponding
zero-photon probabilities $P_{0,\rm max}$ are also given. From the
table it can be seen that for values of $V_R=0.3$ currently achievable
in integrated optics the best choice is to have two multiplexed
units, which on the other hand, does not lead to a significant
improvement compared to the value of $P_1=\exp(-1)\approx 0.368$
achievable with a single unit. The best performance achievable with
any spatial multiplexer and ideal detectors is $P_{1,\rm max}=0.737$,
in which case the theoretical maximum of $V_R=0.95$ is assumed, and we
need $N_{\rm opt}=16$ multiplexed systems, thus $m_{\rm
  opt}=\log_2N_{\rm opt}=4$ cascaded levels. The data clearly show
that the negative effect of real detectors can be compensated by a
higher $\lambda_{\rm opt}$ even for a detector efficiency as low as
$V_D=0.2$, but the achievable single-photon probability is of course
lower.

\subsection{Storage cavity based multiplexers}\label{ssec:cavity}
\begin{figure}[tb]
\includegraphics[width=\columnwidth]{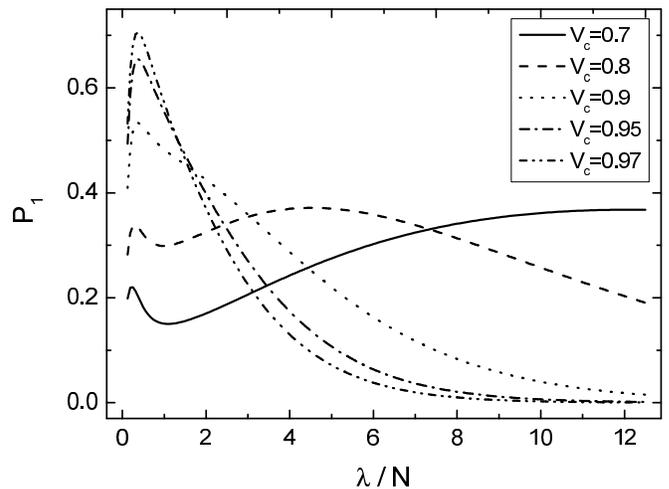}
\caption{\label{fig:tgy:P1LpN}
The single-photon probability $P_1$ plotted against the mean photon pair number per time windows $\lambda/N$ for various storage cavity transmission coefficients $V_c$, considering $N=8$ time windows and assuming ideal detectors and no generic losses ($V_D=V_b=1$).}
\end{figure}
\begin{figure}[tb]
\includegraphics[width=\columnwidth]{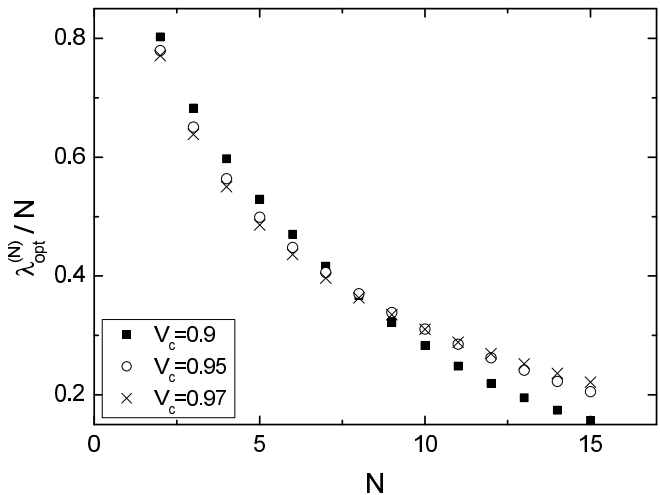}
\caption{\label{fig:tgy:LpNN}
The optimal choice $\lambda_{\rm opt}^{(N)}/N$ of the mean photon pair number per time window as a function of the number of time windows $N$ for various storage cavity transmission coefficients $V_c$, considering $N=8$ time windows and assuming ideal detectors and no generic losses ($V_D=V_b=1$).}
\end{figure}
\begin{figure}[tb]
\includegraphics[width=\columnwidth]{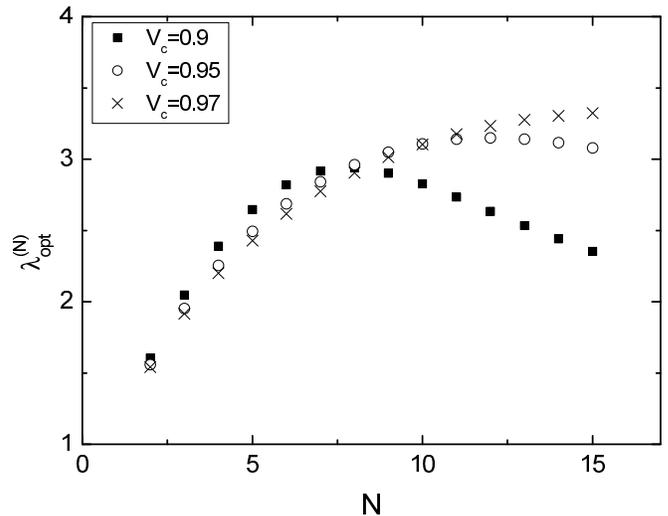}
\caption{\label{fig:tgy:lamN}
The optimal choice $\lambda_{\rm opt}^{(N)}$ of the mean photon pair number plotted against the number of time windows $N$ for various storage cavity transmission coefficients $V_c$, assuming ideal detectors and no generic losses ($V_D=V_b=1$).}
\end{figure}
\begin{figure}[tb]
\includegraphics[width=\columnwidth]{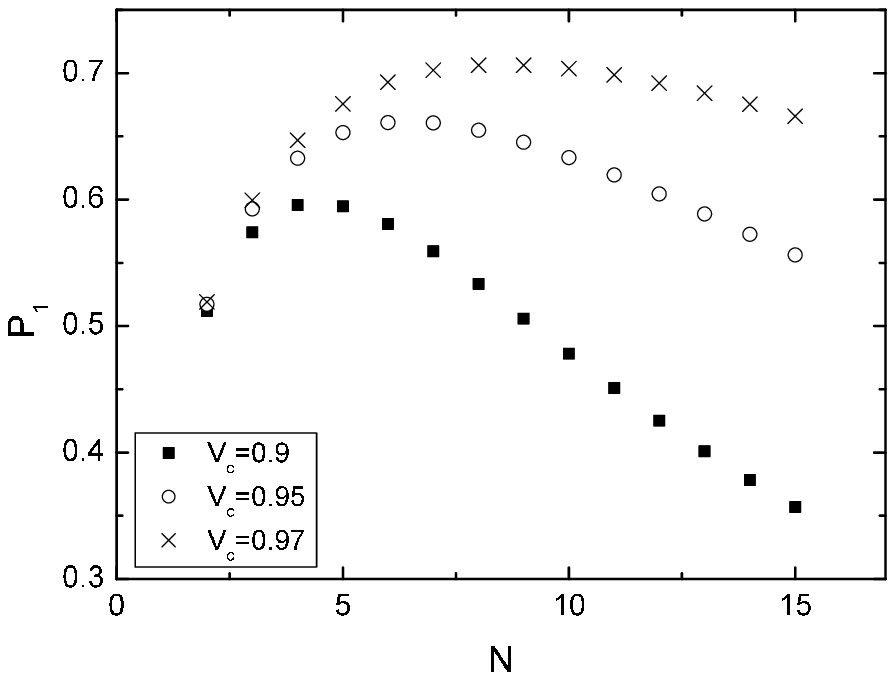}
\caption{\label{fig:tgy:P1N}
The achievable maximal single-photon probability $P_1$ at mean photon pair number $\lambda_{\rm  opt}^{(N)}$ as a function of the number of time windows $N$ for various storage cavity transmission coefficients $V_c$,  assuming ideal detectors and no generic losses ($V_D=V_b=1$).
}
\end{figure}
Before turning our attention to the bulk time multiplexing arrangement proposed in the present paper, let us analyze the storage cavity based multiplexing system first. Throughout this Section we assume ideal detectors ($V_D=1$) and no generic losses ($V_b=1$). In Fig.~\ref{fig:tgy:P1LpN} the dependency of the single-photon probability $P_1$ on $\lambda/N$ is to be seen for various storage cavity transmission coefficients $V_c$  and $N=8$ time windows, the latter being the counterpart of the multiplexed units in the present case. Compared to Fig.~\ref{fig:ter:P1L} displaying the similar relations in the case of spatial multiplexing, the similarity of this dependency is apparent. There is, however, an interesting difference: as the losses increase ($V_c\leq0.8$), a local maximum of $P_1$ appears for small $\lambda/N$. An additional difference is that for smaller losses ($V_c\geq0.9$) the optimal mean photon pair number per time windows corresponding to the maximal single-photon probability decreases instead of increasing with increasing loss. This tendency continues at the mentioned local maxima for bigger losses (that is, for $V_c<0.9$). One can understand this behavior by realizing that in such systems the decrease of the mean photon pair number may also yield improvement in the single-photon probability, as it makes more likely that the photon arrives later, closer to the end of the observation time $T$, and thus it spends less time in the storage cavity, where it is subjected to loss. This effect competes with the increase of single-photon probability due to higher mean photon pair number per time windows, resulting in a local maximum beside the global one for bigger losses.

Figure~\ref{fig:tgy:LpNN} shows the dependence of  optimal $\lambda_{\rm opt}^{(N)}/N$ (corresponding to the maximal values of $P_1$), while Fig.~\ref{fig:tgy:lamN} shows that of  $\lambda_{\rm opt}^{(N)}$ on the number of time windows. In this case $N$ can be any integer (in contrast to the restriction to powers of two in case of spatial multiplexing), hence we use a linear scale instead of a semi-logarithmic one which we use for all the other studied systems in the respective figures. For periodically pumped SPDC sources, $\lambda_{\rm opt}^{(N)}$ corresponds to the mean joint photon pair number of $N$ multiplexed pulses. It appears that in this system, $\lambda_{\rm opt}^{(N)}$ first increases, then starts to decrease with increasing $N$. The decrease of the cavity transmission results in a decrease of the values of $N$ for which $\lambda_{\rm opt}^{(N)}$ grows, while the decrease after the maximum becomes faster. Hence the curves for different $V_c$ intersect. The reason is that due to the losses it is beneficial if the photon gets into the storage cavity as late as possible to spend less time in that lossy environment. In Fig.~\ref{fig:tgy:LpNN} we can observe that, similarly to what we found in the case of the ideal multiplexing system, the value of $\lambda_{\rm opt}^{(N)}/N$ corresponding to the optimum decreases with the increase of the number of time windows $N$. The curves corresponding to different losses intersect again, as for small $N$-s the optimal mean photon pair number of a single time window compensates for the losses, while for a larger number of time windows,  $\lambda_{\rm opt}^{(N)}/N$ is lower to decrease the time the photon spends in the storage cavity.

In Fig.~\ref{fig:tgy:P1N} we have plotted the maximal single-photon probabilities $P_1$ as a function of the number of time windows, for different losses. As it can be seen, this function has a maximum at a given $N_{\rm opt}$. As losses increase, this optimal choice of the number of time windows decreases as for a higher loss it is more likely that the photon is lost in the storage cavity if it spends more time there, which deteriorates the benefits of multiplexing. We remark here that the value of $V_c=0.97$ is the realistic value corresponding to an implementation of the control of the photons with polarizing beam splitters. In this case the maximal single-photon probability is $P_{1,\max}=0.706$ achieved at $\lambda_{\rm opt}=3.014$ with $N_{\rm opt}=9$ time windows.

\subsection{Bulk time multiplexer}

In what follows we analyze the bulk time multiplexer depicted in Fig.~\ref{fig:ido:model}, proposed by us. We assume the generic transmission coefficient $V_b=1$, and we consider all other kinds of losses discussed in Sec.~\ref{sec:statdesc}. In Tab.~\ref{tab:ido} we list all the particular combinations of transmittivity parameters $V_r$, $V_{r,0}$ and $V_t$ we have analyzed, including the best triple available in state-of-art experiments in line number 4.
\begin{table*}
\caption{\label{tab:ido}
Maximal single-photon probabilities $P_{1,\max}$ and the $m_{\rm opt}=\log_2N_{\rm opt}$ number of delay branches and the $\lambda_{\rm opt}$ at which they can be achieved in the bulk time-multiplexed scheme for various loss parameter combinations. We also list the respective zero photon probabilities $P_{0,\max}$. The third column serves as the legend for the figures of this section.}
\begin{ruledtabular}
\begin{tabular}{lclll|rrll|rrll}
&&&&&\multicolumn{4}{c|}{$V_D=1.0$}&\multicolumn{4}{c}{$V_D=0.2$}\\
No.&Sign in Figs.&$V_r$&	$V_{r,0}$&	$V_t$&$m_{\rm opt}$&	$\lambda_{\rm opt}$&	$P_{1,\max}$&$P_{0,\max}$&$m_{\rm opt}$&	$\lambda_{\rm opt}$&	$P_{1,\max}$&$P_{0,\max}$\\\hline
1.&	$\blacksquare$		&1&		1&		1&15&	11.09&	0.999&	1.5e-5&	15&	44.45&	0.999&	1.4e-4\\
2.&	$\bullet$			&1&		1&		0.95&15&	6.85&	0.956&	0.0439&	15&	33.62&	0.955&	0.0439\\
3.& $\plustimes$		&0.996&	0.97&	0.99&7&	7.00&	0.887&	0.0903&	10&	35.24&	0.843&	0.1341\\
4.&	$\blacktriangle$	&0.996&	0.97&	0.95&7&	6.60&	0.858&	0.1222&	10&	33.27&	0.815&	0.1646\\
5.&	$\circ$				&0.996&	0.97&	0.9&6&	5.21&	0.822&	0.1484&	9&	26.26&	0.781&	0.1890\\
6.&	$\blacktriangledown$&0.98&	0.97&	0.95&6&	5.19&	0.806&	0.1662&	9&	26.50&	0.749&	0.2240\\
7.&	$\times$			&0.97&	0.97&	0.95&5&	4.41&	0.779&	0.1748&	8&	22.75&	0.715&	0.2410\\
8.&	$+$					&0.96&	0.97&	0.95&5&	4.37&	0.755&	0.2021&	8&	22.71&	0.684&	0.2767\\
\end{tabular}
\end{ruledtabular}
\end{table*}

\begin{figure}[tb]
\includegraphics[width=\columnwidth]{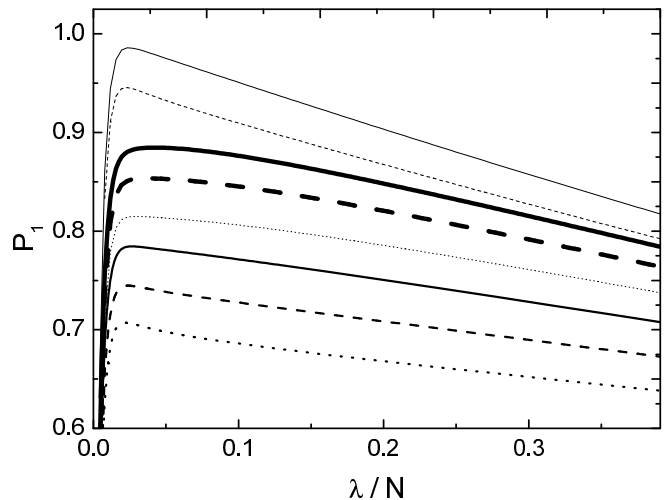}
\caption{\label{fig:ido:P1L}
The single-photon probability $P_1$ as a function of $\lambda/N$ for different combinations of loss parameters, considering a system of $N=256$ time windows, that is, of $m=8$ delay units for the proposed bulk time multiplexer, assuming ideal detectors and no generic losses ($V_D=V_b=1$). From top to bottom the curves presented in the figure correspond to the transmission values  indicated in Tab.~\ref{tab:ido} from 1 to 8.}
\end{figure}

In Fig.~\ref{fig:ido:P1L} we have plotted the single-photon probability $P_1$ as a function of $\lambda/N$ for a system of $N=256$ time windows, that is, of $m=8$ delay units. For a given $N$, and given values of the transmission coefficients, the function has a maximum, thus there exists a value $\lambda_{\rm opt}^{(N)}$ for which the single-photon probability is maximal. This appears to be the case for any $N$. The physics behind it is the same as discussed at the ideal multiplexers. Note that when we decrease any of the three parameters $V_r$, $V_{r,0}$, and $V_t$, the achievable single-photon probability will decrease. We remark that for bigger losses (not shown) the behavior of the function $P_1(\lambda/N)$ will be similar to what we have seen in Fig.~\ref{fig:tgy:P1LpN} for storage cavity based schemes, but the local maxima will appear only for certain combinations of the transmission coefficients. In fact the decrease of the mean photon pair number per time window can only compensate for $V_t$.
\begin{figure}[tb]
\includegraphics[width=\columnwidth]{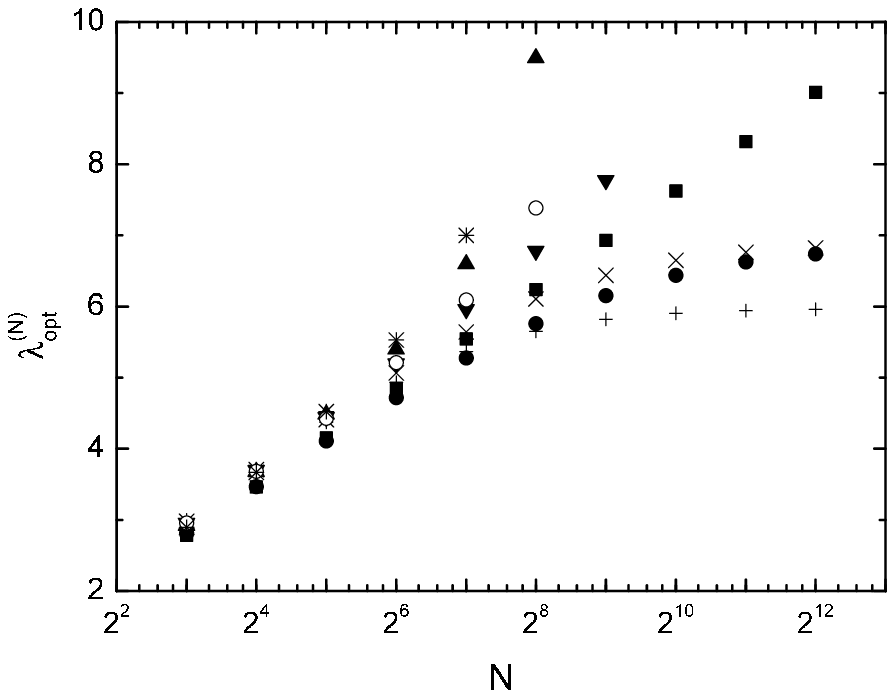}
\caption{\label{fig:ido:lamN}
The optimal choice $\lambda_{\rm opt}^{(N)}$ of the mean photon pair number as a function of the number of time windows $N$ on semi-logarithmic scale for various loss coefficients for the proposed bulk time multiplexer, assuming ideal detectors and no generic losses ($V_D=V_b=1$). The legend of the symbols is in Tab.~\ref{tab:ido}.}
\end{figure}
\begin{figure}[tb]
\includegraphics[width=\columnwidth]{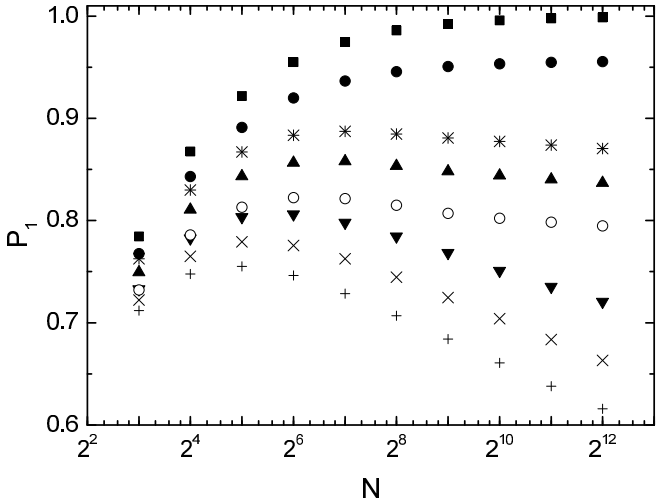}
\caption{\label{fig:ido:P1N}
The achievable maximal single-photon probability $P_1$ at the optimal choice $\lambda_{\rm  opt}^{(N)}$ of the mean photon pair number as a function of the number of time windows $N$ on semi-logarithmic scale for various loss coefficients for the proposed bulk time multiplexer, assuming ideal detectors and no generic losses ($V_D=V_b=1$). The legend of the symbols is in Tab.~\ref{tab:ido}.}
\end{figure}

In Fig.~\ref{fig:ido:lamN} we present the optimal choice $\lambda_{\rm opt}^{(N)}$ of the mean photon pair number as a function of the number of time windows $N$ on semi-logarithmic scale for various loss coefficients presented in Tab.~\ref{tab:ido}. From the curves corresponding to the parameter sets No.~1-2 and 3-4-5 in Tab.~\ref{tab:ido} one
can conclude that increasing the losses proportional to the length of
the delay branches ($V_t'<V_t$) while keeping the coefficients $V_r$ and $V_{r,0}$ (arising from the use or the omission of a delay unit, respectively) constant, the optimal mean photon pair number $\lambda_{\rm opt}^{(N)}$ required for the maximal single-photon probability decreases with the number of time windows. The explanation is similar to the reasoning given in the previous subsection for the other time multiplexing scheme: the decreased mean photon pair number leads to photons generated closer to the end of the observation time $T$, thereby decreasing the necessary delay time and the probability of losing the photon in the medium of the delay system. From curves 4., 6., 7. and 8. we can learn that if $V_t$
and $V_{r,0}$ are fixed, (in particular, $V_t=0.95$, $V_{r,0}=0.97$)
then for $V_r>V_{r,0}$ (that is, if the loss arising from the use of a delay branch is higher than that of its avoiding) the optimal mean photon pair number required for the maximal single-photon probability decreases, while for $V_r<V_{r,0}$ it increases with the number of time windows. This can be expected as the number of activated branches increases with the increase of $\lambda$ resulting in smaller losses for $V_r>V_{r,0}$. On the other hand, in the case of $V_r<V_{r,0}$ the decrease of the mean photon pair number decreases the number of activated branches leading to smaller losses.

In Fig.~\ref{fig:ido:P1N} we have plotted the maximal single-photon probability $P_1$ as a function of the number of time windows, for various loss coefficients. It appears that if there are only propagation losses ($V_r=V_{r,0}=1$) in the system (curves 1-2), the function shows an increasing behavior asymptotically, while if there are delay unit losses ($V_r, V_{r,0}< 1$), these curves also have a maximum. Of course if $V_r=V_{r,0}=1$, there is no disadvantage whatsoever in increasing the number of delay units and at the same time the number of time windows, while accompanying decrease of the size of the time windows the multiphoton probability decreases, which is an advantage. Upon the presence of delay unit losses ($V_r, V_{r,0}< 1$), however, the increase of the number of branches shall increase the zero-photon probability at the output which is the competing disadvantage.

In Tab.~\ref{tab:ido} we have listed the achievable maximal single-photon probabilities $P_{1,\rm max}$ along with the required number of $m_{\rm opt}$ delay branches, determined from the maxima of the curves of Fig.~\ref{fig:ido:P1N}. The corresponding values of $\lambda_{\rm opt}$ are also listed. We have also calculated how these parameters are modified if we have real photodetectors, e.g.\ photomultipliers in the arrangements with a quantum efficiency of $\eta=0.2$, corresponding to $V_D=0.2$. It appears that the achievable maximal single-photon probabilities $P_{1,\rm max}$ do not decrease significantly, only the required mean photon pair number and the number of branches changes in this case.
We calculated $P_{1,\rm max}$ for the best parameters available in the state-of-art experiment (No.~4 in Tab.~\ref{tab:ido}) and for an effective detector such as an avalanche diode ($V_D=0.9$), and obtained 85.4\% at $\lambda_{\rm opt}=6.92$ and $m_{\rm opt}=7$. This single-photon probability is the best that seems to be experimentally realizable nowadays in the analyzed multiplexed periodic single-photon sources.

\section{Conclusions}\label{sec:concl}
We gave an overview of the multiplexed periodic single-photon sources
studied in the literature.  We have suggested a novel time-multiplexed
scheme in bulk optics. Thus far only spatial multiplexing has been
demonstrated experimentally, however, as these schemes require a
relatively large number of components, their scalable realization is
more feasible in integrated optics. If done so, a variety of
problems arise, those related to coupling the input and output fields
to the systems, for instance, which are not present in the case of
bulk optics. Our proposal is the first one for time multiplexing in
bulk optics up to our knowledge, and all its elements are available in
current experiments.

We have introduced a theoretical framework for the statistical
description of all the studied schemes, including the spatial and time
multiplexing ones. We have taken into account all the losses which may
arise in the schemes. The application of this analysis shows that
multiplexing systems can be optimized in order to produce maximal single-photon probability for various sets of loss parameters  by the appropriate choice of the number of multiplexed units of spatial multiplexers or multiplexed time intervals and the input mean photon pair number, and reveals the physical reasons of
the existence of the optimum. We have performed this optimization for
the studied schemes. This may be of use for the optimal design of a
spatial or time multiplexer of this kind. The analysis shows that a
promising single-photon probability of 85\%\ is feasible with the
time-multiplexed scheme in bulk optics we have proposed.

The presented study can serve as a good basis for a design and
realization of an SPDC-based periodic single-photon source, which
would be a necessary device for performing many optical quantum
information processing tasks as well as fundamental quantum optical
experiments.

\begin{acknowledgments}
We thank the support of the National Research Fund of Hungary OTKA (Contract No.\ K83858).
\end{acknowledgments}
%

\end{document}